\documentclass[aps,reprint,twocolumn,superscriptaddress,floatfix]{revtex4-2}

\usepackage{soul}
\usepackage{xcolor}
\usepackage{amsmath,graphicx,mathrsfs}
\usepackage{graphicx}
\usepackage{epstopdf}
\usepackage{mathtools}
\usepackage{upgreek}
\usepackage{bm}
\usepackage{color}
\usepackage{braket}
\usepackage{soul}
\usepackage{lipsum}
\setstcolor{red}

\newcommand{\IN}{\textup{in}}
\newcommand{\OUT}{\textup{out}}
\newcommand{\ADD}{\textup{add}}

\begin{document}

\title{Non-universality of quantum noise in optical amplifiers operating at exceptional points}

\author{L. Simonson}
 
\affiliation{Department of Physics, Michigan Technological University, Houghton, Michigan, 49931, USA}

\affiliation{Henes Center for Quantum Phenomena, Michigan Technological University, Houghton, Michigan, 49931, USA}

\author{S. K. Ozdemir}

\affiliation{Department of Engineering Science and Mechanics, and Materials Research Institute, The Pennsylvania State University, University Park, Pennsylvania 16802, USA}

\author{A. Eisfeld}

\affiliation{Max Planck Institute for the Physics of Complex Systems, N\"othnitzer Strasse 38, 01187 Dresden, Germany}

\author{A. Metelmann}
\email[Corresponding author: ]{anja.metelmann@kit.edu}
\affiliation{Dahlem Center for Complex Quantum Systems and Fachbereich Physik, Freie Universit\"{a}t Berlin, 14195 Berlin, Germany}
\affiliation{Institute for Theory of Condensed Matter, Karlsruhe Institute of Technology, 76131 Karlsruhe, Germany}
\affiliation{Institute for Quantum Materials and Technology, Karlsruhe Institute of Technology, 76344 Eggenstein-Leopoldshafen, Germany}

\author{and R. El-Ganainy}
\email[Corresponding author: ]{ganainy@mtu.edu}

\affiliation{Department of Physics, Michigan Technological University, Houghton, Michigan, 49931, USA}

\affiliation{Henes Center for Quantum Phenomena, Michigan Technological University, Houghton, Michigan, 49931, USA}

\begin{abstract}
The concept of exceptional points-based optical amplifiers (EPOAs) has been recently proposed as a new paradigm for miniaturizing optical amplifiers while simultaneously enhancing their gain-bandwidth product. While the operation of this new family of amplifiers in the classical domain provides a clear advantage, their performance in the quantum domain has not yet been evaluated. Particularly, it is not clear how the quantum noise introduced by vacuum fluctuations will affect their operation. Here, we investigate this problem by considering three archetypal EPOAs structures that rely either on unidirectional coupling, parity-time (PT) symmetry, or particle-hole symmetry for implementing the exceptional point (EP). By using the Heisenberg-Langevin formalism, we calculate the added quantum noise in each of these devices and compare it with that of a quantum-limited amplifier scheme that does not involve any exceptional points. Our analysis reveals several interesting results: most notably that while the quantum noise of certain EPOAs can be comparable to those associated with conventional amplifier systems, in general the noise does not follow a universal scaling as a function of the exceptional point but rather varies from one implementation to another. 
\end{abstract}

\maketitle

\section{Introduction}
Optical amplifiers are the backbone of modern photonics technology. Typical amplifier schemes rely on traveling wave structures which enjoy large bandwidth of operation at the expense of a relatively large footprint. The possibility to shrink the size of optical amplifiers by using optical cavity structures has been demonstrated in a number of studies \cite{Bjorlin2001JQE, Cole2005JQE, Bjorlin2003JQE}. Unfortunately, these cavity-based devices suffer from a limitation imposed by their gain-bandwidth product. In other words, one can increase the gain by sacrificing the bandwidth and vice versa. Recently, our groups introduced the notion of exceptional points (EPs) \cite{Feng2017NPho, El-Ganainy2018NP, Ozdemir2019NM, Miri2019S} based optical amplifiers (EPOAs), which enjoy both a small footprint due to their cavity-based construction and an enhanced gain-bandwidth product enabled by the presence of EPs \cite{Zhong2020PRApp}. As a result, an EPOA operates at a larger gain and bandwidth than their counterpart cavity-based devices that do not possess an EP.  Interestingly, in this early work, it was found that the scaling of the gain-bandwidth product is not universal but rather depends on how the system supporting the EPs is implemented. For an amplifier made of PT-symmetric \cite{Ruter2010NP, Peng2014NP, Hodaei2014Sci} coupled resonators, the values of the gain and bandwidth can be decoupled from each other, allowing for the building of devices with arbitrary parameters that are limited only by the fabrication constraints. On the other hand, for devices based on chiral EPs \cite{Wiersig2014PRL,Hashemi2021APL} and exceptional surfaces \cite{Zhong2019PRL, Zhong2019OL, Zhong2021PRR, SoleymaniNatComm2022}, the gain-bandwidth product is a function of the respective EP order (i.e., the number of the coalescing eigenvalues and eigenvectors). This early work considered only operation in the classical domain and hence quantum noise was neglected. Given the current intense activities in quantum photonics technology \cite{Wang2020NPho}, it is of interest to study the operation of EPOAs in the quantum domain and evaluate their performance in the presence of quantum noise. While earlier studies have investigated the role of quantum noise in specific PT symmetry arrangements \cite{Yoo2011PRA, Agarwal2012PRA}, here we focus on the performance of EP-based amplifiers by considering various, realistic optical schemes that can be used to realize EPOAs. These include configurations that feature either unidirectional coupling, PT symmetry, or particle-hole symmetry. Several approaches can be used to investigate the effect of quantum fluctuations in these systems \cite{Weiss1999Book} such as the Lindblad master equation or the quantum Langevin equations. Here, we adopted the quantum Langevin technique for calculating the input-output relation, redand we use an effective model that involves one parameter for modeling each loss channel and one parameter for modeling the gain. The physical origin of the quantum noise is attributed to the coupling between the amplifier structure and these open channels (vacuum fluctuations in the loss channels and noise arising from light-matter coupling under pumping conditions to provide gain - see for instance \cite{gardiner00}).\\

\begin{figure} [!t]
	\includegraphics[width=3.4in]{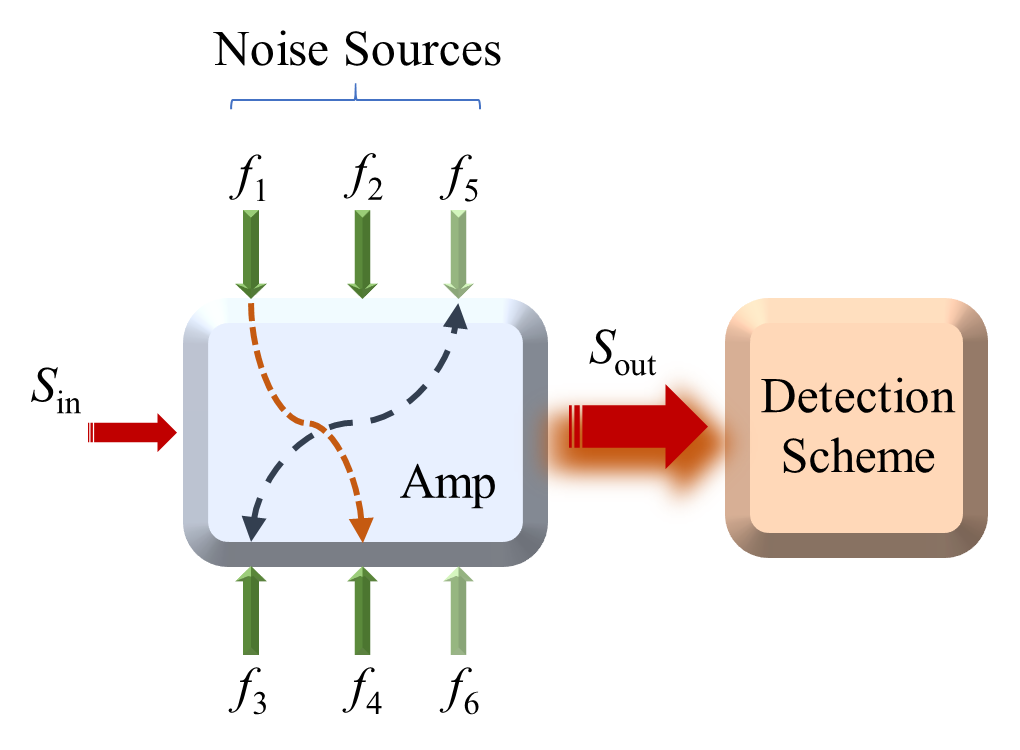}
	\caption{A schematic of an optical amplification and detection system. The input signal $S_{\IN}$ is boosted at the output $S_{\OUT}$ with the addition of extra noise (indicated by the blurring of the output arrow). The added noise value depends on the noise contribution from each independent open channel and their mutual interactions. To illustrate the main goal of this work, a hypothetical amplifier configuration with hypothetical open channels $f_{1-4}$ is depicted. Tuning the system to operate at an EP may introduce new open channels, say $f_{5,6}$, and/or feedback between some of these channels as illustrated by the dashed blue and red arrows for bidirectional and unidirectional feedback scenarios. However, the quantum noise of each individual channel is not affected since this is dictated by the Heisenberg uncertainty principle. These effects combined will consequently change the added noise at the output. Possible detection schemes include coherent or direct detection. In this work, we evaluate the added noise for some practical implementations of EPOAs for coherent detection, which is widely used in communication networks.}
	\label{Fig_Schematic}
\end{figure}

Before we proceed, it is beneficial to first recap some of the main results in the quantum theory of phase-insensitive amplifiers \cite{Caves1982PRD}, and in doing so, outline the physics of the problem. The frequency domain input-output field relation for a phase insensitive amplifier is described by the canonical expression for the amplifier response \cite{Caves1982PRD}:

\begin{equation} \label{Eq.Ino-out}
	\hat{S}_{\OUT}[\omega] = \mathcal{G}[\omega] \left( \hat{S}_{\IN}[\omega] + \hat{\mathcal{N}}[\omega] \right),
\end{equation}

where $\hat{S}_{\IN}$ and $\hat{S}_{\OUT}$  are the bosonic annihilation operators at the input/output channels, and  
$\hat{\mathcal{N}}[\omega]=\sum_{m}^{} c_m[\omega] \hat{\eta}_m[\omega] + \sum_{n}^{} d_n[\omega] \hat{\xi}_n^{\dagger}[\omega]$ is the total noise operator, with $\hat{\eta}_{m}[\omega]$ and 
$\hat{\xi}_{n}[\omega]$ being the Gaussian white noise operators associated with individual loss/gain channels \cite{Clerk2010RMP}, indicated by the subscripts $m,n$. They obey the statistics  $\braket{\hat{\eta}^{\dagger}_{m}[\omega]\hat{\eta}_{m'}[\omega']}=
\braket{\hat{\xi}_n^{\dagger}[\omega] \hat{\xi}_{n'}[\omega']} =0$,
$\braket{\hat{\eta}_{m}[\omega] \hat{\eta}_{m'}^{\dagger}[\omega']} = 2\pi \delta(\omega+\omega')\delta_{m,m'} $ and $\braket{\hat{\xi}_{n}[\omega] \hat{\xi}_{n'}^{\dagger}[\omega']} = 2\pi \delta(\omega + \omega') \delta_{n,n'}$, with the expectation values involving cross terms, i.e. both $\hat{\eta}$ and $\hat{\xi}$, vanishing. In the above, the expectation values are evaluated with respect to the vacuum state since thermal noise can be neglected  at optical frequencies. In writing the above correlation relations, we used the convention $\hat{A}[\omega] \sim \int \hat{A}(t) e^{i\omega t} dt$ and $\hat{A}^{\dagger}[\omega] \sim \int \hat{A}^{\dagger}(t) e^{i\omega t} dt$, i.e. $\hat{A}^{\dagger}[\omega]=(\hat{A}[-\omega])^{\dagger}$, for any operator $\hat{A}$. These definitions imply that operating at any frequency $\omega_o$ corresponds to $\omega=\omega_o$ and $\omega'=-\omega_o$. In what follows, we will focus on the case when $\omega_o$ is the resonant frequency of the microresonators. Finally, the amplitude amplification factor $\mathcal{G}[\omega]$, and the coefficients $c_m[\omega]$ and $d_n[\omega]$, are complex functions whose actual values depend on the amplifier structure. By imposing the bosonic commutation relation on $\hat{S}_{\OUT}$ in Eq. (\ref{Eq.Ino-out}), we obtain the constraint: 

\begin{equation} \label{Eq.Comm_Cond}
	1 +  \sum_{m}^{}|c_m[\omega]|^2-\sum_{n}^{}|d_n[\omega]|^2 = \frac{1}{G[\omega]},
\end{equation}

where $G[\omega]=|\mathcal{G}[\omega]|^2$ is the power amplification factor. The above relation is generic and applies to any linear, phase insensitive optical amplifier. While the impact of the amplifier noise on the input signal depends on the amplifier's design, the measured noise at the output is a function of both the amplifier's structure and the particular light detection schemes (see Fig. \ref{Fig_Schematic}) which can be coherent (heterodyne/homodyne) or incoherent (direct detection). The former requires complex setups and measures information encoded in the electric field (which is often used with very high-speed communication networks) while the latter requires simpler setups and measures the power without providing any information about the phase. In what follows, we will focus on homodyne detection and discuss direct detection briefly in Appendix C.
Under this condition, the added noise is given by $\bar{n}_{\ADD} \equiv \mathcal{S}[\omega_o]$, where 
$\mathcal{S} [\omega] \equiv \int d\omega'/4\pi \;  \braket{\{\hat{\mathcal{N}}[\omega],\hat{\mathcal{N}}^{\dagger}[\omega'] \}}$ is the symmetrized noise spectral density \cite{Caves1982PRD, Clerk2010RMP} and $\{\}$ denotes anticommutationIntuitively, $S[\omega]$ quantifies the sum of the uncertainties associated with measuring the two quadratures of the electric field \cite{Caves1982PRD}. From the mathematical definition of $\mathcal{S} [\omega]$, it follows that:

\begin{equation} \label{Eq.Noise}
        \mathcal{S} [\omega] =   \frac{1}{2}\left(\sum_{m}^{}|c_m[\omega]|^2+\sum_{n}^{}|d_n [\omega]|^2\right).
\end{equation}

\begin{figure} [!t]
	\includegraphics[width=3in]{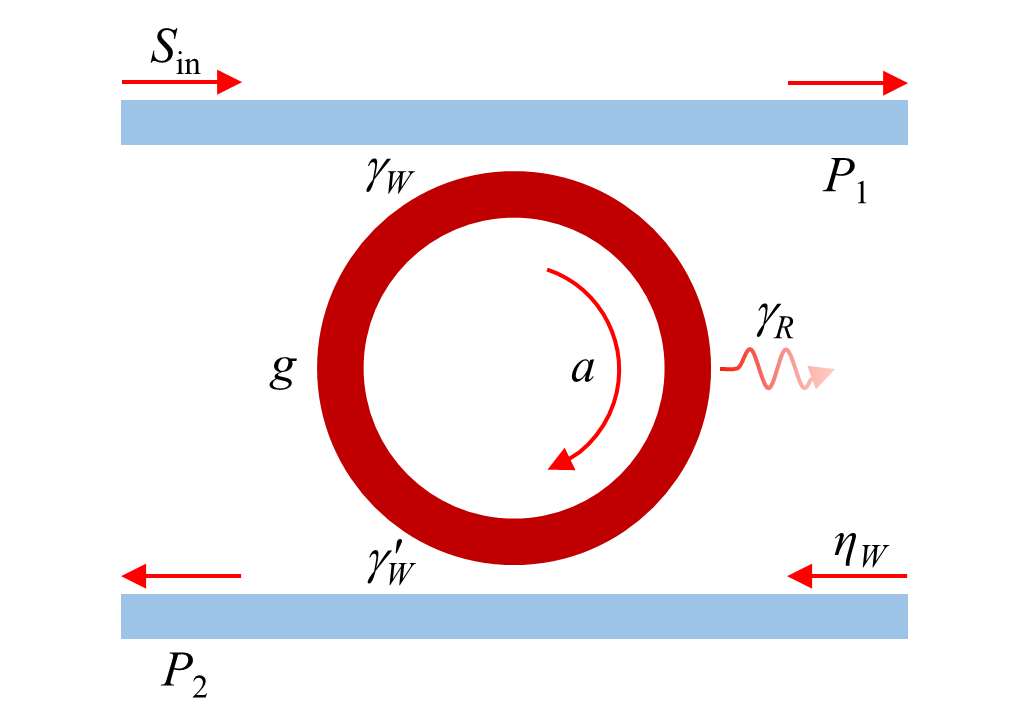}
	\caption{An optical amplifier made of a microring resonator (gain element) with evanescently coupled waveguides serving as input/output channels. In the ideal case of negligible back reflection, only one optical mode (here the CW wave) contributes to the amplification process. When $\gamma'_W=0$, the output from port $P_1$ exhibits a minimum quantum noise of $\bar{n}_{\ADD} \approx \frac{1}{2}$. On the other hand, when $\gamma'_W=\gamma_W$, the output from port $P_2$ will feature a purely Lorentzian amplification but at the same time will have a larger quantum noise of $\bar{n}_{\ADD} \approx \frac{3}{2}$.}
	\label{Fig.Linear_Amp}
\end{figure} 

If there is only one input noise channel associated with the applied gain (i.e. in the absence of any $c$ terms and by retaining only one $d$ term), such as for the OA shown in Fig. \ref{Fig.Linear_Amp} under the condition $\gamma'_W=0$, and if the output is collected from port $P_1$ (see Fig. \ref{Fig.Linear_Amp}), Eqs. (\ref{Eq.Comm_Cond}) and (\ref{Eq.Noise}) give $\bar{n}_{\ADD}=  \frac{G_o-1}{2G_o}$, where $G_o \equiv G[\omega_o]$ is the power gain at resonance. In the limit of large $G_o$, this yields $\bar{n}_{\ADD} \approx \frac{1}{2}$. This value represents the quantum limit of linear, phase-insensitive  OAs \cite{Caves1982PRD}. On the other hand, Eqs. (\ref{Eq.Comm_Cond}) and (\ref{Eq.Noise}) do not impose such a bound when the system involves more than one noise channel. For instance, the amplifier system shown in Fig. \ref{Fig.Linear_Amp} will have a noise floor of $\bar{n}_{\ADD} \approx \frac{3}{2}$ when $\gamma'_W=\gamma_W$ and the output is collected from port $P_2$. By recalling that the formation of an EP often involves the fine tuning of Hermitian and non-Hermitian parameters, EPOAs will typically have several noise channels associated with the various non-Hermitian degrees of freedom. In addition, the formation of an EP can introduce coupling and/or coherent feedback between some of these channels. Thus, while the presence of an EP does not change the fundamental noise properties in each open channel (this is dictated by the Heisenberg uncertainty principle), it may introduce interference effects between the different channels and alter the total output noise value (see Fig. \ref{Fig_Schematic} for a schematic illustration of this discussion). Hence, it is not a priori clear whether EPOAs are quantum-limited or not. \\

In this work, we investigate this question by considering various, realistic implementation schemes of EPOAs featuring a second order EP. To analyze the noise properties of these systems, we employ the Heisenberg-Langevin formalism. Before we present the details of our calculations, we first summarize the main results. Our analysis reveals that the quantum noise in EPOAs does not follow a universal scaling with the order of the EP but rather varies widely depending on the actual photonic implementation. For instance, the noise in an amplifier featuring a chiral EP can be very different from that associated with an implementation based on PT-symmetry. This is a rather surprising result given that the noise enhancement factor in laser systems is expected to follow a universal scaling behavior according to the Petermann factor. Importantly, we find that for some implementations, the noise performance of an EPOA can be comparable to that of conventional amplifiers based on diabolic points (DPs) while simultaneously providing an advantage in terms of the gain-bandwidth scaling.    
 
\begin{figure} [!t]
	\includegraphics[width=3in]{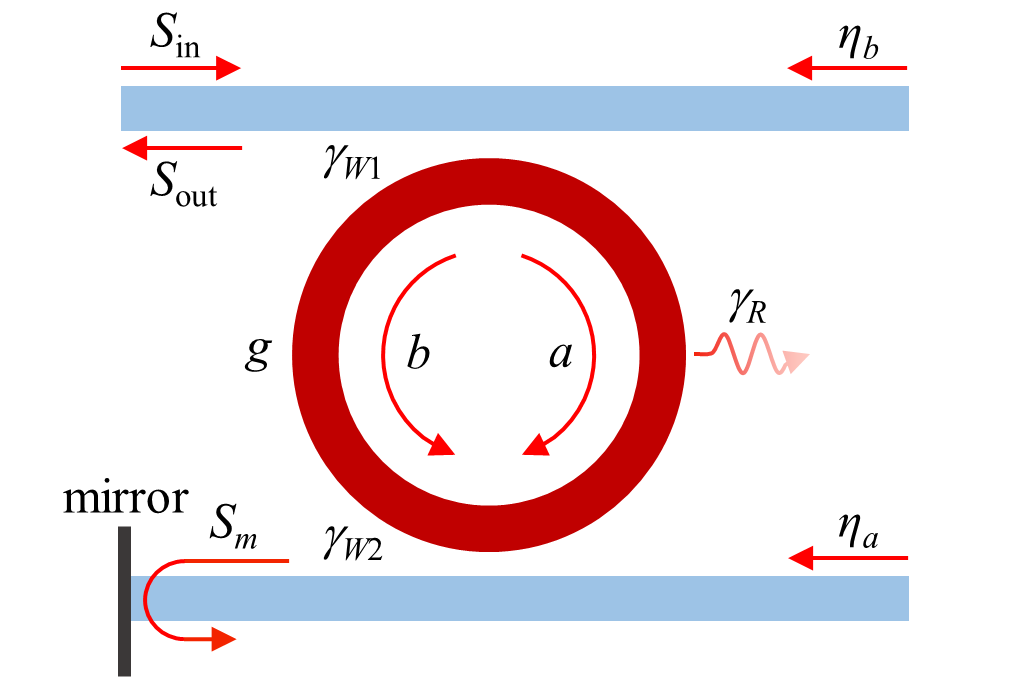}
	\caption{An optical amplifier configuration that implements a chiral EP via unidirectional coupling (for details, see \cite{Zhong2020PRApp}). The input/output channels as well as the noise sources are also depicted on the figure.}
	\label{Fig.Mirror_Amp}
\end{figure} 

\section{Results}
In what follows, we evaluate the added quantum noise associated with several different implementations of optical amplifiers having an exceptional point of order two. In all of our calculations, we employ Eq. (\ref{Eq.Ino-out}) together with the definitions of $\hat{\mathcal{N}}$ and $\bar{n}_{\ADD}$ to calculate the quantum noise.\\

\textit{Amplifiers with chiral EPs and two coupling channels:} We start our analysis by considering an EPOA based on unidirectional coupling as shown in Fig. \ref{Fig.Mirror_Amp}. Achieving asymmetric coupling can be accomplished by a variety of techniques (see for instance \cite{Wiersig2016PRA, Hashemi2021APL}). Here we focus on a simple configuration that employs the coupling between a microring resonator and a waveguide with an end mirror \cite{Zhong2019PRL, Zhong2020PRApp, Zhong2019OL, Zhong2021PRR} as shown in Fig. \ref{Fig.Mirror_Amp}. For this structure, the Heisenberg-Langevin equations, expressed in a frame rotating with the resonant frequency of the bare resonator, take the form:

\begin{subequations} \label{Eq.Langevin_Chiral}
	\begin{align}
		\begin{split}
			\frac{d}{dt}\hat{a} &= \Gamma \hat{a} + \sqrt{2\gamma_{W1}} \hat{S}_{\IN} + \sqrt{2\gamma_R} \hat{\chi}_a  \\
			&\quad{} + \sqrt{2\gamma_{W2}} \hat{\eta}_a + \sqrt{2g}\hat{\xi}_a^{\dagger}, \label{Eq.Langevin_ChiralA}
		\end{split}
		\\ \hat{S}_m &= \hat{\eta}_a-\sqrt{2\gamma_{W2}} \hat{a}, \label{Eq.Langevin_ChiralB} \\
		\begin{split}
			\frac{d}{dt}\hat{b} &=\Gamma \hat{b} + \sqrt{2\gamma_{W2}} e^{i\phi} \hat{S}_m + \sqrt{2\gamma_R} \hat{\chi}_b \\
			&\quad{} + \sqrt{2\gamma_{W1}}\hat{\eta}_b + \sqrt{2g}\hat{\xi}_b^{\dagger}, \label{Eq.Langevin_ChiralC}
		\end{split}
		\\ \hat{S}_{\OUT} &= \hat{\eta}_b-\sqrt{2\gamma_{W1}} \hat{b}, \label{Eq.Langevin_ChiralD}
	\end{align}
\end{subequations}

where $\hat{a}$ and $\hat{b}$ are the annihilation operators associated with the clockwise (CW) and counterclockwise (CCW) modes of the microring resonators, respectively, and $\Gamma = g - \gamma_{W1} -\gamma_{W2} - \gamma_R$. Here $\gamma_{W1,2}$  denote the coupling to the two waveguides and $\gamma_R$ accounts for intrinsic losses in the resonator, as illustrated in Fig. \ref{Fig.Mirror_Amp}. Additionally, $\phi$ represents the phase acquired by the mode upon traveling twice along the waveguide and reflecting from the mirror (see Fig. \ref{Fig.Mirror_Amp}). Crucially, the two loss mechanisms described by $\gamma_{W1,2}$ and $\gamma_R$ two independent noise channels for each mode. In addition, the gain process comes with it's own noise channel.\\
	
In the absence of a drive and noise, the system can be described using the non-Hermitian Hamiltonian $H = \bigl( \begin{smallmatrix} i\Gamma & 0\\ -2ie^{i\phi}\gamma_{W2} & i\Gamma \end{smallmatrix}\bigr)$, which in the bases  $e^{-i \lambda t}$ (i.e. when the stationary solution is expressed as $\hat{x}(t)=\hat{X}e^{-i\lambda t}$, with $x=a,b$) has the eigenvalues $\lambda_{1,2} = i\Gamma$, i.e. it exhibits a second order chiral EP (the term chiral here refers to the propagation direction inside the ring, CW or CCW, and not to the chirality associated with encircling EPs \cite{Heiss2001EUD}).

From Eq. (\ref{Eq.Langevin_Chiral}), it is straightforward to show that the power amplification factor of the above amplifier is given by $G_o  =  16 \gamma_{W1}^2 \gamma_{W2}^2/\Gamma^4 $, with the high gain limit obtained when the system approaches the lasing threshold ($g = \gamma_{W1} + \gamma_{W2} + \gamma_R$) from below. As has been demonstrated in \cite{Zhong2020PRApp}, this scheme has an enhanced gain-bandwidth scaling compared to a standard cavity-based OA that does not exhibit an EP. On the other hand, the EPOA described via Eqs. (\ref{Eq.Langevin_Chiral}) has two additional  external noise channels compared to a standard OA with one waveguide. Under steady state conditions, we find (see Appendix A for a detailed derivation): 

\begin{equation} \label{Eq.Chiral_Noise}
\begin{aligned}
	\bar{n}_{\ADD} = \frac{1}{2}\bigg[\frac{\gamma_{W2}+\gamma_R+g}{\gamma_{W1}}  + \frac{\Gamma^2}{4\gamma_{W2}^2}\bigg(1+\frac{\gamma_{W2}+\gamma_R+g}{\gamma_{W1}}\bigg)\\+\frac{\Gamma^4}{16\gamma^2_{W1}\gamma^2_{W2}}\bigg].
\end{aligned}
\end{equation}

In the high gain limit of $g \to \gamma_{W1} + \gamma_{W2}$, and under the realistic assumption of $\gamma_R \ll \gamma_{W1,2}$ (both conditions taken together imply that $\frac{\Gamma}{\gamma_{W1,2}} \to 0$), Eq. (\ref{Eq.Chiral_Noise}) reduces to $\bar{n}_{\ADD} = \frac{1}{2} + \frac{\gamma_{W2}}{\gamma_{W1}}$. This expression shows that the quantum noise can be decreased by choosing $\gamma_{W2} \ll \gamma_{W1}$. However, this will also decrease the gain dramatically the output signal from the port indicated on the figure vanishes altogether for $\gamma_{W2}=0$). In fact, the gain can be expressed as a function of the noise factor according to: 

\begin{equation} \label{Eq.Chirl_Gain_noise}
	 G_o = \frac{16\gamma_{W1}^4 }{\gamma_R^4} \left( \bar{n}_{\ADD}-\frac{1}{2} \right)^2.
\end{equation}

Note that in this system, $\bar{n}_{\ADD} \geq \frac{1}{2}$. Of practical importance is the case when $\gamma_{W1}=\gamma_{W2}$. Under this condition, the gain can attain a large value of $G_o=\frac{16\gamma_{W1}^4}{\gamma_R^4}$. However, the noise is now given by $\bar{n}_{\ADD}=\frac{3}{2}$, i.e. exactly similar to the DP-based amplifier shown in Fig. \ref{Fig.Linear_Amp} with two waveguides. This may indicate that the extra noise is mainly due to the additional waveguide channel rather than an intrinsic feature of EPs themselves. Importantly, however, despite sharing the same noise values,
the EPOA of Fig. \ref{Fig.Mirror_Amp} enjoys a better gain-bandwidth scaling as compared to the DP amplifier. 
\\

\begin{figure} [!t]
	\includegraphics[width=3.5in]{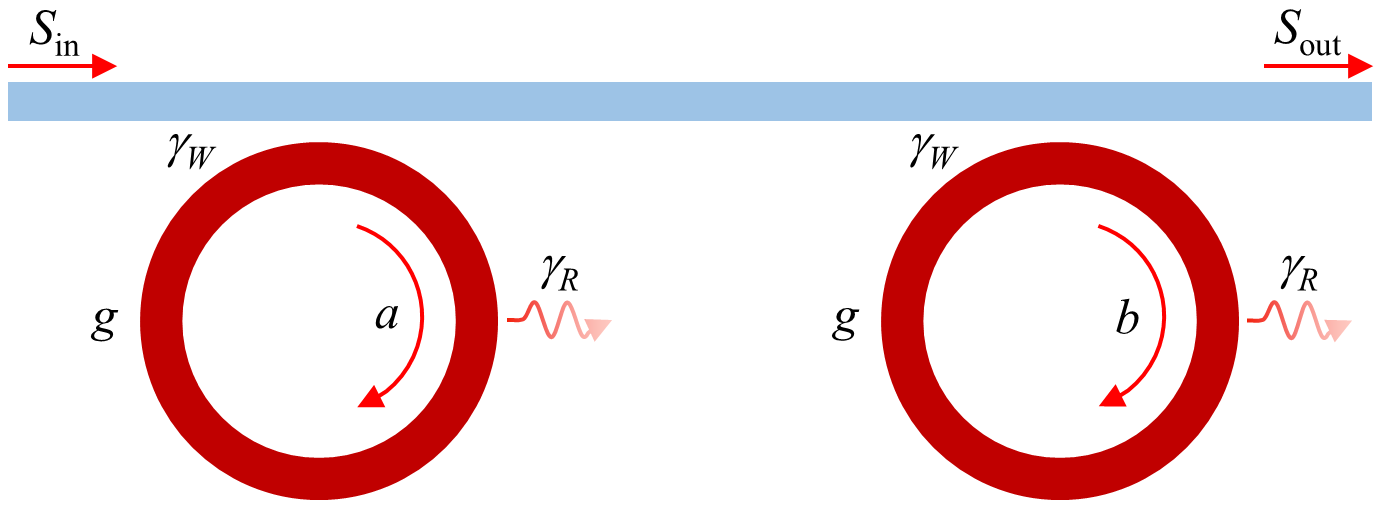}
	\caption{A series of two identical optical amplifiers with CW modes $a$ and $b$ coupled through a common waveguide. Both resonators have the same gain $g$ with identical coupling coefficients $\gamma_W$ and radiation loss $\gamma_R$.}
	\label{Fig.Linear_Two_Amp}
\end{figure}

To test whether the extra noise is an intrinsic feature of EPs themselves or is due to the additional waveguide, we now consider the geometry shown in Fig. \ref{Fig.Linear_Two_Amp}, where two identical microring resonators are coupled through a common waveguide. By considering the subspace of CW modes and the input/output channel depicted on the figure, we can obtain the system's response as before and show that it features an EP. In this case, we find that the power gain at resonance is given by $G_o = \frac{16\gamma_W^2(g-\gamma_R)^2}{(g-\gamma_W-\gamma_R)^4}$, and the added noise is $\bar{n}_{\ADD}=\frac{1}{2}$ in the large gain limit. This favorable noise scaling, however, is contrasted by a spectral response that features a superposition between Lorentzian and super Lorentzian lineshapes which may suppress the enhancement in the gain-bandwidth product predicted for EPOAs. Thus, while the introduction of an EP in the linear spectrum of an OA does not change the fundamental noise properties of the individual channels, the coherent feedback and interference effects arising due to the presence of the EP can lead to different output noise values compared to those obtained in the DP case.\\

Finally, we remark that in the above analysis we considered a perfectly reflecting mirror for implementing the EP. In Appendix B, we show that considering the more realistic scenario of a partially reflecting mirror does not change the added noise in the high gain limit. This is mainly because the finite reflectivity of the mirror affects the amplification and noise factors in the same manner.\\

\begin{figure} [!t]
	\includegraphics[width=2.6in]{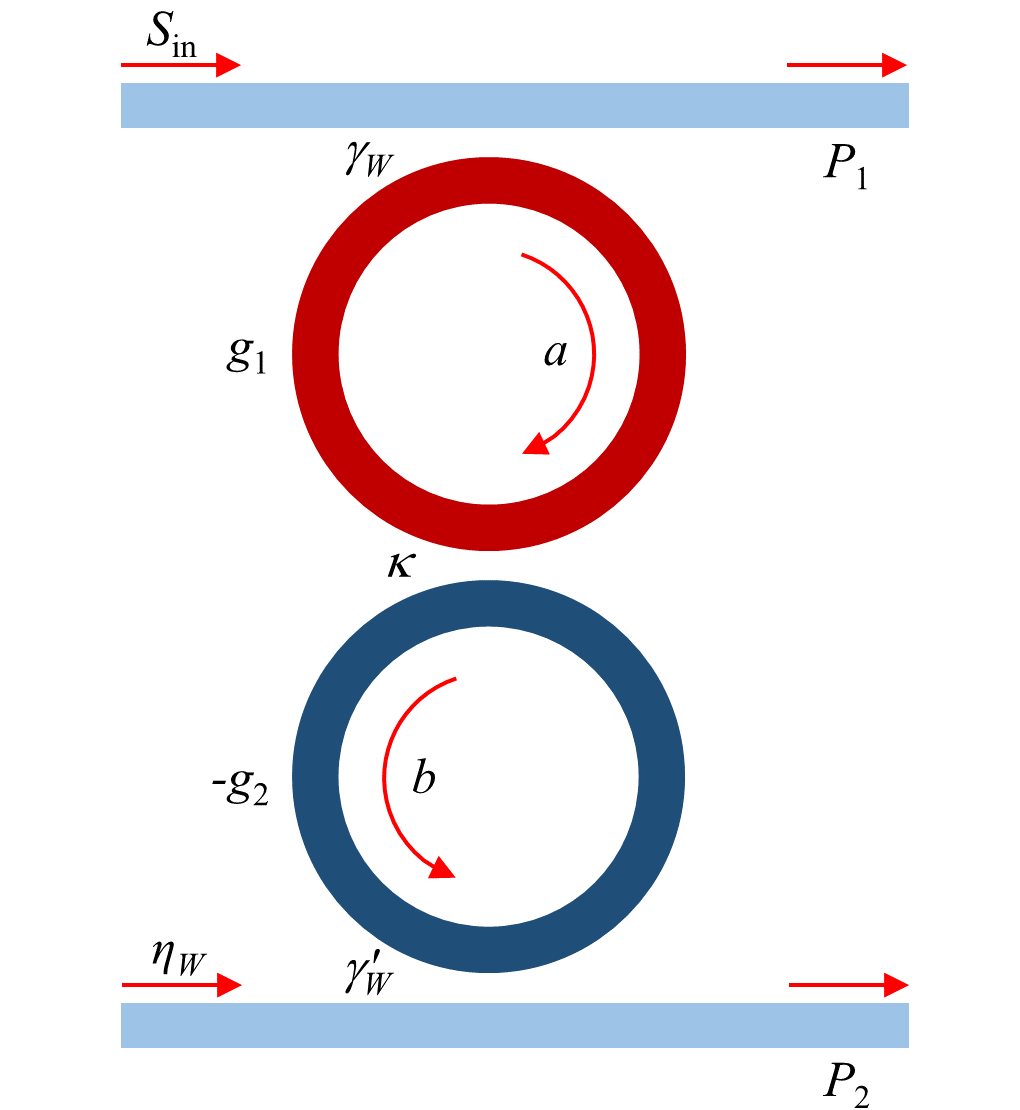}
	 \caption{A PT-based optical amplifier. The two microrings are coupled to each other with a coupling coefficient $\kappa$ and also to identical waveguides with coupling coefficients $\gamma_W$ and $\gamma'_W$. The top ring exhibits a gain $g_1$ while $-g_2$ represents a loss in the bottom ring. In our analysis, when $\gamma'_W=\gamma_W$, the output is collected from port $P_2$. On the other hand, when $\gamma'_W=0$, $P_1$ is the output port.}
	\label{Fig.PT_Amp}  
\end{figure}

\textit{PT-symmetric amplifiers with two coupling channels:} Next, we consider  different implementation of EPs that relies on PT-symmetric arrangements as shown in Fig. \ref{Fig.PT_Amp}. For the input/output channels indicated on the figure, the  Heisenberg-Langevin equations associated with the relevant optical modes of the microring resonators read:

\begin{subequations} \label{Eq.Langevin_PT}
\begin{align}
 \frac{d}{dt} \hat a  =&  \;   -   \left(  \gamma_{W}   - g_{1}  \right) \hat a  - i \kappa \hat b 
                   + \sqrt{ 2\gamma_{W}} \hat S_{ \IN}      
                   + \sqrt{2 g_{1}} \hat \xi_{a}^{\dag}, \\
\frac{d}{dt} \hat b  =&  \;   - \left(\gamma'_{W}   + g_{2}  \right) \hat b  -i \kappa \hat a 
                 + \sqrt{2 \gamma'_{W}} \hat \eta_{W}    
                 + \sqrt{2 g_{2}} \hat \eta_{b }.
\end{align}
\end{subequations}

Here, $\gamma_W$, $\gamma'_W$ and $\kappa$ denote the coupling rate between the resonators and their respective waveguides and the coupling coefficient between the two microring resonators, whereas $g_1>0$ is the gain factor associated with first resonator. On the other hand, $g_2>0$ is an extra loss applied to the second resonator in order to tune the operating point and achieve pseudo-PT symmetry (i.e. PT symmetry up to an additional symmetric gain or loss) when $\gamma'_W=\gamma_W$ and $g_{1} = g_{2}$. Here, we neglected intrinsic losses and assumed resonant excitation. In what follows, we take $\gamma'_W=\gamma_W$. Note that this amplifier architecture is exposed to three noise channels due to the coupling to the waveguide and the gain and loss channels of the resonators. Thus, we obtain one noise channel less than the amplifier described via Eqs. (\ref{Eq.Langevin_Chiral}). In the absence of driving, and in the classical limit, the non-Hermitian Hamiltonian associated with Eqs. (\ref{Eq.Langevin_PT}) is given by $H = \bigl( \begin{smallmatrix} -i(\gamma_W - g_1) & \kappa \\ \kappa & -i(\gamma_W + g_2) \end{smallmatrix}\bigr)$ and its eigenvalues in the bases $e^{-i \lambda t}$ read:

\begin{align} \label{Eq.PT2WGEigenvalue}
\lambda_{1,2} =     -i(\gamma_{W} + g_{-}) \pm \sqrt{ \kappa^2 - g_{+}^2 },
\end{align}
where $g_{\pm} = (g_2 \pm g_1)/2$. Clearly, the spectrum  of $H$ in this case exhibits an EP when $\kappa = g_{+}$. Under an external excitation $\hat{S}_{\IN}$ and for output channel $P_2$, we obtain the following noise expression:

\begin{align}\label{Eq.AddedNoise2waveguides}
\bar n_{\ADD}  =  \frac{1}{2} \left(1 + \frac{1}{G_{o}} -   \frac{2 \Gamma_1}{\gamma_{W}} + \sqrt{G_{o}}  \frac{\Gamma_1^2}{\gamma_{W}^2} \right) ,
\end{align}
with $\Gamma_1 = \gamma_{W} + g_{-}$. As has been noted before \cite{Zhong2020PRApp}, this amplifier has an interesting feature in the classical domain: its bandwidth ($B_{EP} = 2 F \Gamma_1 $ with $F = \sqrt{\sqrt{2} - 1}$), and resonant gain ($G_{o} =  \frac{ 4 \gamma_{W}^2 g_{+}^2}{\Gamma_1^4}$) are decoupled from one another when it operates in the regime $g_{-} > 0$. Therefore, one can achieve high gain by increasing $g_{+}$ by increasing both $g_{1,2}$ without crossing the lasing threshold while keeping the bandwidth constant by fixing $\Gamma_1$. This however comes at the expense of an increased quantum noise given by Eq. (\ref{Eq.AddedNoise2waveguides}) far above the quantum limit. This unusual scaling for the added quantum noise can be understood by noting that the device shown in Fig. \ref{Fig.PT_Amp} is a multi-port structure. Hence, there are various choices for configuring the input and output channels. If a particular configuration is not optimal in the sense that it does not provide the maximum possible signal amplification, then it is possible that the amplification factor for a specific noise channel exceeds that of the signal. It is exactly this effect that leads to the large noise predicted by Eq. (\ref{Eq.AddedNoise2waveguides}) in the high gain limit.

Interestingly, this unfavorable noise scaling can be circumvented by considering the parameter regime of $ g_{-}<0$. 
However, this condition leads to an operation regime that resembles that of a DP amplifier where the large gain limit coincides with the lasing threshold at $\Gamma_1 = 0$. Given that the bandwidth scales with $\Gamma_1$, it is evident that in this regime,  the gain-bandwidth coupling is reintroduced. The upside is that in this same regime of $ g_{-} < 0$, we have $\Gamma_1 / \gamma_W < 1$, which, for some parameter range, may lead to noise suppression. To demonstrate this, we first set $r\equiv \frac{2g_{+}}{\Gamma_1} > 0$ or equivalently $\Gamma_1/ \gamma_W = r/\sqrt{G_o}$. The noise and bandwidth expression of the PT amplifier then take the form:

\begin{subequations} \label{Eq.PTNoise2WG}
\begin{align}
 \bar n_{\ADD} =&   \frac{1}{2} \left[1 + \frac{1}{G_o}  +   \frac{r}{\sqrt{G_o}} \left( r  - 2   \right) \right]  , \\
   B_{EP} &= 2 r F   \frac{ \gamma_W}{\sqrt{G_o}}.
\end{align}
\end{subequations}

\begin{figure*} 
  \centering\includegraphics[width=0.99\textwidth]{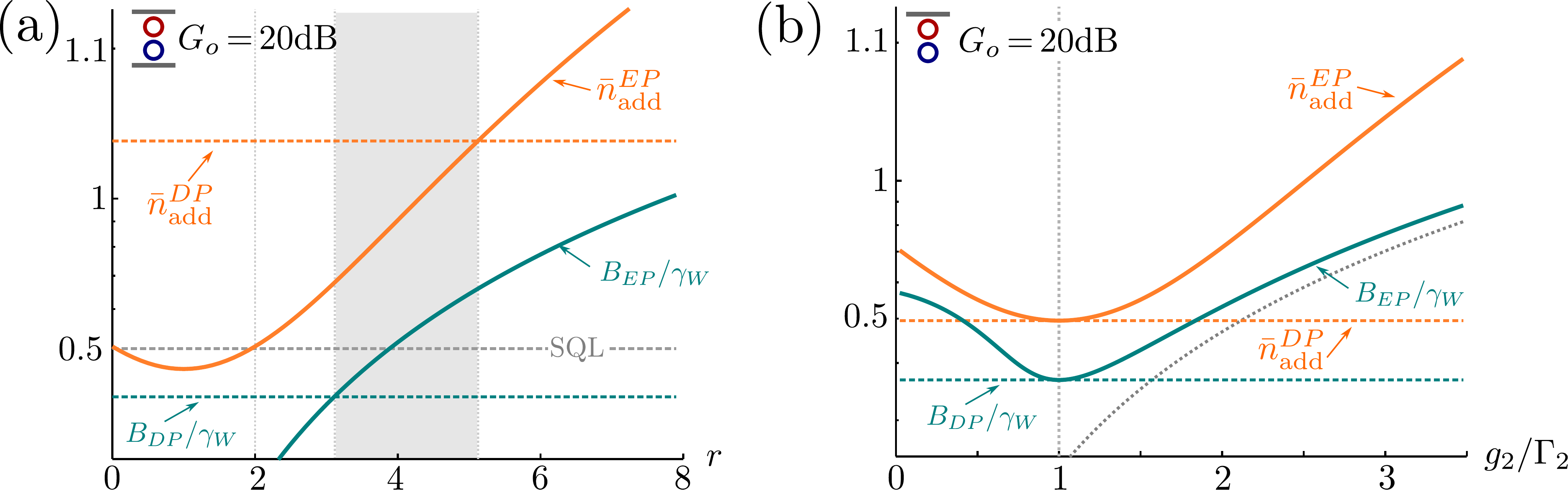} 
 	\caption{Characteristics of PT-like optical amplifiers operated at an exceptional point. The dashed lines denote the corresponding results for a DP-amplifier. (a) Added noise $\bar n_{\rm add}^{EP}$ and bandwidth $B_{EP}$ for the PT-like architecture with two waveguides as a function of the factor $r = \sqrt{G_{o}} \Gamma_1 /\gamma_{W}  $. The grey area denotes the regime where the PT-like architecture outperforms the DP-amplifier in terms of reduced noise and enhanced bandwidth. Note that we only plot the interesting domain of $B_{EP}/\gamma_W$ since this quantity vanishes at $r=0$. (b) Added noise $\bar n_{\rm add}^{EP}$ and bandwidth $B_{EP}$ for the PT-like architecture with one waveguide as a function of the loss rate $g_2/\Gamma_2$.}
 	\label{Fig.:BandwidthNoise}
\end{figure*}

Thus, compared to a DP-amplifier with gain-bandwidth product of $4 \gamma_{W}$, we have $B_{EP}/B_{DP} =r F /2$.
Fig. \ref{Fig.:BandwidthNoise}(a) depicts the added noise and bandwidth as a function of the parameter $r$.
When $1-\sqrt{1-1/\sqrt{G_o}}<r<1+\sqrt{1-1/\sqrt{G_o}}$ (which becomes $0<r<2$ in the high gain limit), the added noise is below the quantum limit of half a photon. However, in that same domain, the corresponding bandwidth is far below the bandwidth of
the DP-amplifier. When $r\sim 2$ in the aforementioned high gain limit, the added noise equals that of a DP-amplifier with a single waveguide, having significantly lower added noise as the DP-amplifier with two waveguides, but with 
bandwidth reduced by a factor $F$. For $r > 1+\sqrt{1+1/\sqrt{G_o}}$, the bandwidth continuously increases together with the added noise. 
The upshot of this analysis is that, when compared to a two-channel DP-amplifier, the PT amplifier with two waveguides can exhibit improved gain-bandwidth scaling for $r > 2/F $, and less noise for $r < 1 + \sqrt{2} G_{o}^{1/4}$. The intersection of these two intervals thus define the optimal operating regime as shown by the shaded gray area in Fig. \ref{Fig.:BandwidthNoise}. 

\textit{PT-symmetric amplifiers with one coupling channel:} It is instructive at this point to also consider a PT amplifier geometry with only one waveguide channel. This situation corresponds to Fig. \ref{Fig.PT_Amp} with $\gamma'_W=0$. The output channel in that case is port $P_1$. The Langevin equations for this system are:

\begin{subequations}
\begin{align} \label{Eq.Langevin_PT1WG}
 \frac{d}{dt} \hat a  =&    - \left(  \gamma_W  - g_{1}  \right) \hat a  - i \kappa \hat b
                   + \sqrt{ 2\gamma_W} \hat S_{\IN}    
                   + \sqrt{2 g_{1}} \hat \xi_{a}^{\dag}  , \\
\frac{d}{dt} \hat b =&   -    g_{2} \hat b  -i \kappa \hat a  
                 + \sqrt{2 g_{2}} \hat \eta_{b}.
\end{align}
\end{subequations}

Here the number of noise channels is only two, as opposed to three for the two waveguide setup. The noise channels are again arising due to the gain and loss channels of the resonators and the coupling to the waveguide (neglecting intrinsic losses for simplicity). For this system, the non-Hermitian Hamiltonian now reads as $H = \bigl( \begin{smallmatrix} -i(\gamma_W - g_1) & \kappa \\ \kappa & -i g_2 \end{smallmatrix}\bigr)$. In the classical domain, the eigenvalues of the structure in the bases $e^{-i \lambda t}$ become:

\begin{align} \label{Eq.PT1WGEigenvalue}
\lambda_{1,2} =  -i\left(\frac{\gamma_W}{2} + g_{-}\right)  \pm   \sqrt{  \kappa ^2 - \left(\frac{\gamma_W}{2} - g_{+}\right)^2 } .
\end{align}
The spectrum exhibits an EP for $\kappa = \left|\gamma_{W}/2 -g_{+}\right|$ and the lasing threshold coincides with $g_{-} > -  \gamma_{W}/2 $.

At this EP point, and below the lasing threshold, the power gain as a function of frequency detuning from the resonant frequency is given by

       \begin{align}  \label{Eq.PTGainSingleWG}
       G[\Delta \omega] =   1    
                   + \frac{     4\gamma_{W}    g_1 \; \Delta \omega^2         
                        }{ \left(  \Delta  \omega^2 + \Gamma_2^{2} \right)^2 }     
                   +   \frac{  4 \gamma_{W}   g_2 \left( 
                                 \gamma_{W}     g_2     -  \Gamma_2^{2}
                                \right)
                   }
                   {  \left(  \Delta \omega^2 + \Gamma_2^{2} \right)^2 } ,
\end{align} 
with $\Gamma_2 = \frac{\gamma_{W}}{2} +   g_{-} $. This expression reflects the fact that the output features an interference between three different trajectories. The first term arises due to direct transmission path from the input channel to the output port. The second and third terms can be rearranged into the sum of two Lorentzian functions and a super-Lorentzian of order two, which is a characteristic noted before for systems with second order EPs under general excitation conditions \cite{Khanbekyan2020PRR, SoleymaniNatComm2022,  Zhong2021PRR}. As before, in the regime of $ g_{-} > 0$, this system realizes an amplifier with decoupled  gain and bandwidth. However, a constant gain-bandwidth product comes again with the price of having additional noise. The amplifier cannot be quantum-limited as the noise is amplified more strongly than the signal just as for the two waveguide architecture. Thus, we focus on the regime where $ - \gamma_{W}/2 < g_{-} < 0$, with a coupling between the gain and bandwidth. Here, an enhancement of the
bandwidth is still expected due to the presence of a double pole in the super-Lorentzian term's denominator while noise contributions should scale with the same gain factor. More specifically, the added noise reads:

\begin{align} \label{Eq.NoisePT1WG}
   \bar n_{\ADD} =   
                     \left(  \frac{   g_2 -  2 \Gamma_2 }{\gamma_{W}} + \frac{1}{2}   \right)  \left(  1 +  \frac{   1 }{  \sqrt{G_{o} }   }    \right)^2 
                   +  \frac{1}{\sqrt{G_{o} }} + \frac{1}{ G_{o}  }    ,   
\end{align} 
where again, we used $G_{o} = G[\Delta \omega =0]$. In the large gain limit, which is achieved as  $\Gamma_2 \rightarrow 0$, the added noise becomes:

\begin{align} \label{Eq.NoiseLargeGainPT1WG}
   \lim_{\Gamma_2 \rightarrow 0} \bar n_{\ADD} =  \frac{1}{2} + \frac{ g_{2}}{\gamma_{W}} ,      
\end{align}
which shows that the system can approach the quantum limit when $g_{2} \ll \gamma_{W}$.  
  Fig.~\ref{Fig.:BandwidthNoise}(b) plots the noise and bandwidth of this amplifier as a function of $g_2/\Gamma_2$ when the resonant gain value is $G_o=20$dB. As expected from Eq. (\ref{Eq.NoisePT1WG}), the added noise approaches the value of $\sim 0.7$ in the limit $g_{2} \rightarrow 0$. Interestingly, increasing the value of $g_{2}$ decreases the added noise until $g_{2} = \Gamma_2$, where the noise reaches its minimum value. An expression for the minimum noise can be obtained by using the relations $g_{2} = \Gamma_2$, $\Delta\omega=0$ in Eq. (\ref{Eq.PTGainSingleWG}), solving for $\gamma_W/\Gamma_2$, and substituting back into Eq. (\ref{Eq.NoisePT1WG}). Doing so gives:
  
\begin{align} \label{Eq.MinNoisePT1WG}
   \lim_{g_{2} = \Gamma_2} \bar n_{\ADD}   =    \frac{1}{2}    \left(  1    - \frac{  1 }{   G_{o}  }  \right)    ,            
\end{align}
which has the same form as the added noise of the DP-amplifier with a single waveguide.  Importantly, we also note that at this same point, the bandwidth attains its minimum value and the expression for the gain takes the form: 

\begin{align} \label{Eq.Gain@MinNoisePT1WG}
  \lim_{g_2 = \Gamma_2} G[\Delta \omega] =   
                   1         
                   +   \frac{  4  \gamma_W   \left(    \gamma_W        -      g_2   \right)
                   }
                   {  \left(\Delta   \omega^2 +   g_2^2  \right) } .
\end{align}

Note that the super-Lorentzian term does not exist at this point and the gain expression scales similar to that of a DP-amplifier. This analysis shows that
a PT-like amplifier with one waveguide is only quantum-limited when it behaves similar to single-pole DP-amplifier. Operating away from this point leads to bandwidth enhancement, but at the expense of additional noise.\\

\textit{Amplifiers with EP implementation via particle-hole symmetry:} For completeness, we have also studied an EP implementation that features particle-hole symmetry \cite{Ryu2002PRL, Kawabata2019NC, Yang2017PRA, Ge2017PRA}. It is described by the non-Hermitian Hamiltonian $H= \bigl(\begin{smallmatrix} \Delta \omega + i(g - \gamma_W) & i\kappa \\ i\kappa & - \Delta \omega + i(g - \gamma_W) \end{smallmatrix} \bigl)$, where here we explicitly retain the dependence on the detuning $\Delta \omega$. Note that the Hamiltonian $H$ anti-commutes with the PT operator, hence it is also called anti-PT-symmetric \cite{Ge2013PRA, Zhang2020PRL, Ge2017PRA}. The eigenvalues of $H$, as written in the bases $e^{-i \lambda t}$, are given by $\lambda_{1,2} = i(g-\gamma_W) \pm \sqrt{\Delta \omega^2-\kappa^2}$, which exhibits an EP when $\kappa=\Delta\omega$. Implementing the above Hamiltonian requires dissipative coupling \cite{Gentry2014OL}, which is expected to add more noise terms, but for clarity we neglect this here and consider only noise contributions similar to that of Eqs. (\ref{Eq.Langevin_PT}). At the EP, the gain and noise of this system are given by

\begin{equation} \label{Eq.Diss_Gain_NoiseOP}
		G[\Delta \omega] = \frac{4\gamma_W^2[\Delta \omega^2 +(g - \gamma_W)^2]}{(g - \gamma_W)^4} ,
\end{equation}

and

\begin{equation} \label{Eq.Diss_Noise}
\begin{aligned}
	\bar{n}_{\ADD} = \frac{1}{2}\bigg[\frac{\Delta \omega^2}{\Delta \omega^2 + (g-\gamma_W)^2} \\ +  \frac{g}{\gamma_W} \left(1 + \frac{\Delta \omega^2}{\Delta \omega^2 + (g-\gamma_W)^2}\right)\bigg] .
	\end{aligned}
\end{equation}

In the high gain limit ($g \to \gamma_W$), the above expression reduces to $\bar{n}_{\ADD} = \frac{3}{2}$, which is not quantum-limited. In reality, it is expected that the actual noise level of such a system will be even larger than this value when accounting for the loss channels introduced by implementing the dissipative coupling.

While the above analysis does not constitute a general proof that EP-based optical amplifiers with enhanced gain-bandwidth products are not quantum-limited, it shows clearly that practically feasible devices will indeed have noise above the minimum possible value for conventional amplifiers. Interestingly, the level of added quantum noise correlates with the scaling enhancement in the gain-bandwidth product. Particularly, for the configuration depicted in Fig. \ref{Fig.Mirror_Amp}, where the aforementioned scaling is enhanced but the values of the gain and bandwidth remain interconnected, the added noise increases but remains finite. On the other hand, for a two waveguide PT-symmetric arrangement (as in Fig. \ref{Fig.PT_Amp}) where the gain and bandwidth are completely decoupled in the case when $g_->0$, the added noise value blows up. As we explained earlier, this observation can be understood by noting that in the first case, the large gain limit is achieved for a finite value of the optical gain coefficient $g$, while in the second, it is approached for $g_+ \to \infty$.

\section{Discussion and Conclusion}

In conclusion, we have calculated the quantum noise associated with optical amplifiers operating at EPs for various device implementations by using the Heisenberg-Langevin approach. Our analysis shows that the added noise due to vacuum fluctuations can vary dramatically between two different devices corresponding to different implementations for the same EP. In other words, the noise values depend on the device topology. For instance, an amplifier structure that features a chiral EP with an enhanced gain-bandwidth product will exhibit a quantum noise larger than but close to that of an amplifier operating at a DP. On the other hand, a PT-symmetric amplifier operating in a regime where the gain and bandwidth are totally decoupled will have a divergent noise value in the large gain limit. These results indicate that there is a trade-off between the improved device performance (defined by the enhanced gain-bandwidth product) and the added quantum noise. Importantly, however, the quantum noise does not scale linearly with the order of the EP. In fact, for certain device geometries, the quantum noise may even saturate as the order of the EP is increased infinitely. To demonstrate this effect, we consider the structure depicted in Fig. \ref{Fig_High_Order_Amp}. It is straightforward to see that a similar device made of $M$ rings will exhibit an EP of order $M$. An input signal from the leftmost waveguide will cross every ring before it enters the next stage, i.e. the device features a amplifier made of $M$ cascaded stages, with the total output given by: 

\begin{figure} [!t]
	\includegraphics[width=3.4in]{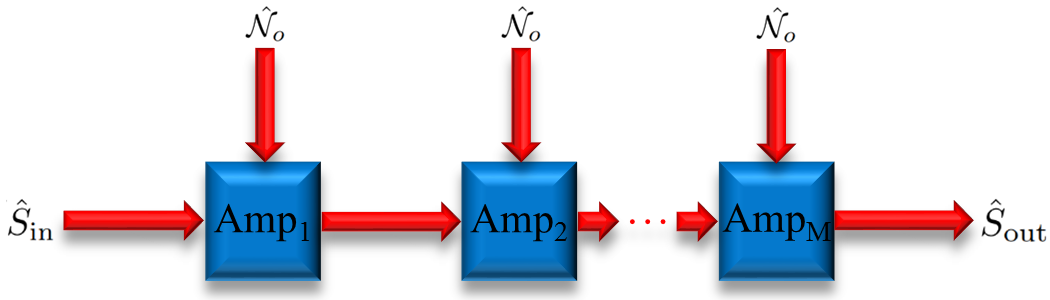}
	\caption{A cascaded optical amplifier configuration with an EP of order $M$. An input signal $\hat{S}_{\IN}$ from  the leftmost waveguide enters amplifier $\text{Amp}_1$ that has quantum noise $\hat{\mathcal{N}_o}$. The output of this amplifier is then fed into a second waveguide and then into a second amplifier that also has quantum noise $\hat{\mathcal{N}_o}$, with this process then repeating for $M$ amplifiers. For large values of gain, the noise of $\text{Amp}_1$ will dominate the quantum noise in $\hat{S}_{\OUT}$.}
	\label{Fig_High_Order_Amp}
\end{figure} 

\begin{equation} \label{Eq.Ino-out_2}
\hat{S}_{\OUT}=\mathcal{G}^M \left(\hat{S}_{\IN} +\hat{\mathcal{N}_o} \sum_{m=0}^{M-1} \mathcal{G}^{-m} \right),
\end{equation}
where as before $\hat{S}_{\IN}$ and $\hat{S}_{\OUT}$ are the bosonic annihilation operators at the input and output channels respectively, $\mathcal{G}$ is the amplitude amplification factor, and $\hat{\mathcal{N}_o}$ is the quantum noise operator associated with each stage (see Fig. \ref{Fig_High_Order_Amp}). For a large value of $\mathcal{G}$, only the first term in the summation (corresponding to $m=0$) will dominate (in fact, for an infinite series, the contribution of the first term exceeds that of all the other terms combined when $\mathcal{G} > 2$) and thus only this term will contribute to total added quantum noise. From our earlier discussion about the DP-based optical amplifier made of one ring resonator and two waveguides, it is clear that the first amplification stage will thus contribute a noise value of $\bar{n}_{\ADD}=\frac{3}{2}$. The upper limit on the noise due to the multistage amplification can be obtained by taking $\lim_{M\to\infty}$. In this case, the summation reduces to $\frac{\mathcal{\mathcal{G}}}{\mathcal{G}-1}$. For $\mathcal{G}=10$, the maximum noise for a device with infinite number of cascaded stages increases only by a factor of $\sim 0.11$ while for $\mathcal{G}=100$, this factor becomes $\sim 0.0101$. These predictions can also be understood in light of the recent work on the linear response theory of resonant non-Hermitian systems \cite{Hashemi2022NC}. Finally, we note that while considering a direct detection scheme will affect the measured noise levels, it does not have a significant impact on the comparison between EP and DP-based amplifiers (see Appendix C). \\

These results, aside from their technical importance for device design, raise some fundamental questions regarding the noise consideration in non-Hermitian photonic systems operating at EPs. Particularly, in the context of laser theory, it was shown that non-Hermitian effects lead to enhanced linewidth close to the lasing threshold as quantified by the Petermann factor \cite{Petermann1979QE}. This description, however, fails at EPs (as it predicts infinite linewidth) and also under highly nonlinear conditions \cite{Amelio2022PRA}. Consequently, this begs the following questions: does the linewidth of an EP-based laser device scale with the EP's order, or does it also depend on the details of the implementation? Does the Petermann factor indeed limit the sensitivity of an EP-based sensor \cite{Wiersig2020PR} regardless of how the sensor is built \cite{Langbein2020PRA, Lau2018NC, Zhang2019PRL, Wang2020NC, Wiersig2020PRA}, or can one cleverly design some sensing devices that circumvent the otherwise predicted bound on sensitivity? The answers to these questions are of fundamental importance for designing next generation photonic devices.

\section*{Appendix A: Derivation of added noise}

While the derivation of $\bar{n}_{\ADD}$ for various amplifier structures is straightforward, here we present the detailed derivation for one example, which we take to be Eq. (\ref{Eq.Chiral_Noise}), for the benefit of the reader. Evaluating the noise starts with solving Eqs. (\ref{Eq.Langevin_Chiral}) that describe the structure in Fig. \ref{Fig.Mirror_Amp} under steady state conditions to express the output field in terms of the input signal. This is achieved by first solving for  $\hat{a}$ (describing CW mode) in Eq. (\ref{Eq.Langevin_ChiralA}):

\begin{equation} \label{Eq.aChiral}
\begin{aligned}
    \hat{a} = -\frac{\sqrt{2\gamma_{W1}}}{\Gamma}\bigg(\hat{S}_{\IN} + \sqrt{\frac{\gamma_R}{\gamma_{W1}}} \hat{\chi}_a \\ + \sqrt{\frac{\gamma_{W2}}{\gamma_{W1}}} \hat{\eta}_a + \sqrt{\frac{g}{\gamma_{W1}}} \hat{\xi}_a^{\dagger}\bigg),
\end{aligned}
\end{equation}

and then substitute into Eq. (\ref{Eq.Langevin_ChiralB}) to arrive at:

\begin{equation} \label{Eq.SmChiral}
\begin{aligned}
    \hat{S}_m = \hat{\eta}_a + \frac{2\sqrt{\gamma_{W1}\gamma_{W2}}}{\Gamma}\bigg(\hat{S}_{\IN} + \sqrt{\frac{\gamma_R}{\gamma_{W1}}}\hat{\chi}_a \\ + \sqrt{\frac{\gamma_{W2}}{\gamma_{W1}}}\hat{\eta}_a
    + \sqrt{\frac{g}{\gamma_{W1}}}\hat{\xi}_a^{\dagger}\bigg).
\end{aligned}
\end{equation}

Repeating the same calculation for the CCW mode by substituting Eq. (\ref{Eq.SmChiral}) into Eq. (\ref{Eq.Langevin_ChiralC}) gives:

\begin{equation} \label{Eq.bChiral}
\begin{aligned}
    \hat{b} = -\frac{1}{\Gamma}\bigg[\frac{2\sqrt{2\gamma_{W1}}\gamma_{W2}e^{i\phi}}{\Gamma}\bigg(\hat{S}_{\IN} + \sqrt{\frac{\gamma_R}{\gamma_{W1}}}\hat{\chi}_a \\ + \sqrt{\frac{\gamma_{W2}}{\gamma_{W1}}}\hat{\eta}_a
    + \sqrt{\frac{g}{\gamma_{W1}}}\hat{\xi}_a^{\dagger}\bigg)  + \sqrt{2\gamma_{W2}}e^{i\phi} \hat{\eta}_a \\ + \sqrt{2\gamma_R}\hat{\chi}_b + \sqrt{2\gamma_{W1}}\hat{\eta}_b + \sqrt{2g}\hat{\xi}_b^\dagger \bigg].
\end{aligned}
\end{equation}

By substituting Eq. (\ref{Eq.bChiral}) into Eq. (\ref{Eq.Langevin_ChiralD}), we obtain the following expression for the output:

\begin{equation} \label{Eq.SoutChiral}
\begin{aligned}
    \hat{S}_{\OUT} =  \frac{4\gamma_{W1}\gamma_{W2} e^{i\phi}}{\Gamma^2}\bigg[\hat{S}_{\IN} +\frac{\Gamma^2}{4\gamma_{W1}\gamma_{W2} e^{i\phi}}\hat{\eta}_b + \sqrt{\frac{\gamma_R}{\gamma_{W1}}}\hat{\chi}_a \\ + \sqrt{\frac{\gamma_{W2}}{\gamma_{W1}}}\hat{\eta}_a + \sqrt{\frac{g}{\gamma_{W1}}}\hat{\xi}_a^{\dagger}+\frac{\Gamma}{2\gamma_{W2}e^{i\phi}}\bigg(\sqrt{\frac{\gamma_R}{\gamma_{W1}}}\hat{\chi}_b \\ + \sqrt{\frac{\gamma_{W2}}{\gamma_{W1}}}e^{i\phi}\hat{\eta}_a + \hat{\eta}_b + \sqrt{\frac{g}{\gamma_{W1}}}\hat{\xi}_b^{\dagger}\bigg)\bigg].
\end{aligned}
\end{equation}

From the definition of the noise operator in Eq. (\ref{Eq.Ino-out}), we find:

\begin{equation} \label{Eq.NoiseOpChiral}
\begin{aligned}
    \hat{\mathcal{N}} =\frac{\Gamma^2}{4\gamma_{W1}\gamma_{W2} e^{i\phi}}\hat{\eta}_b+ \sqrt{\frac{\gamma_R}{\gamma_{W1}}}\hat{\chi}_a + \sqrt{\frac{\gamma_{W2}}{\gamma_{W1}}}\hat{\eta}_a + \sqrt{\frac{g}{\gamma_{W1}}}\hat{\xi}_a^{\dagger}\\+\frac{\Gamma}{2\gamma_{W2}e^{i\phi}}\bigg(\sqrt{\frac{\gamma_R}{\gamma_{W1}}}\hat{\chi}_b + \sqrt{\frac{\gamma_{W2}}{\gamma_{W1}}}e^{i\phi}\hat{\eta}_a + \hat{\eta}_b + \sqrt{\frac{g}{\gamma_{W1}}}\hat{\xi}_b^{\dagger}\bigg).
\end{aligned}
\end{equation}

From this expression, the coefficients $c_m$ and $d_n$ are identified and substituted back into Eq. (\ref{Eq.Noise}) and we obtain the result in Eq. (\ref{Eq.Chiral_Noise}).\\

An important point here is that writing Eq. (4) in a frame rotating with the resonant frequency of the resonator corresponds to using $\bar{n}_{\ADD} \equiv \mathcal{S}[\omega_o=0]$. Of course, one can work in a non-rotating frame and use the actual value of $\omega_o$ to arrive at the same results. \\

\section*{Appendix B: Impact of finite mirror reflectivity}

Here, we examine the impact of the finite mirror reflectivity on the noise associated with the amplifier structure depicted in Fig. \ref{Fig.Mirror_Amp}. We start by noting that Eqs. (\ref{Eq.Langevin_Chiral}) mostly remain the same except for Eq. (\ref{Eq.Langevin_ChiralC}) which is now modified according to:

\begin{equation} \label{Eq.Langevin_ChiralC_r}
\begin{aligned}
    \frac{d}{dt}\hat{b} &=\Gamma \hat{b} + \sqrt{2\gamma_{W2}} r_m e^{i\phi} \hat{S}_m + \sqrt{2\gamma_R} \hat{\chi}_b \\
	&\quad{} + \sqrt{2\gamma_{W1}}\hat{\eta}_b + \sqrt{2\gamma_{W2}} t_m e^{i\theta} \hat{\eta}_m + \sqrt{2g}\hat{\xi}_b^{\dagger}, 
\end{aligned}
\end{equation}

where $r_m$ and $t_m$ are the magnitudes of the field reflection and field transmission coefficients across the mirror. In addition, $\theta$ is the phase of the transmission coefficient. The new noise term $\hat{\eta}_m$ is associated with the newly opened noise channel to the left of the mirror. By following the same algebraic steps as before, we arrive at:

\begin{equation} \label{Eq.NoiseOpChiral_r}
\begin{aligned}
    \hat{\mathcal{N}} =\frac{\Gamma^2}{4\gamma_{W1}\gamma_{W2} r_m e^{i\phi}}\hat{\eta}_b+ \sqrt{\frac{\gamma_R}{\gamma_{W1}}}\hat{\chi}_a + \sqrt{\frac{\gamma_{W2}}{\gamma_{W1}}}\hat{\eta}_a \\ + \sqrt{\frac{g}{\gamma_{W1}}}\hat{\xi}_a^{\dagger}+\frac{\Gamma}{2\gamma_{W2} r_m e^{i\phi}}\bigg(\sqrt{\frac{\gamma_R}{\gamma_{W1}}}\hat{\chi}_b + \sqrt{\frac{\gamma_{W2}}{\gamma_{W1}}} r_m e^{i\phi}\hat{\eta}_a \\ + \hat{\eta}_b + \sqrt{\frac{\gamma_{W2}}{\gamma_{W1}}} t_m e^{i\theta} \hat{\eta}_m + \sqrt{\frac{g}{\gamma_{W1}}}\hat{\xi}_b^{\dagger}\bigg),
\end{aligned}
\end{equation}

which in turn gives:

\begin{equation} \label{Eq.Chiral_Noise_r}
\begin{aligned}
	\bar{n}_{\ADD} = \frac{1}{2}\bigg[\frac{\gamma_{W2}+\gamma_R+g}{\gamma_{W1}} + \frac{\Gamma^2}{4\gamma_{W2}^2 r_m^2}\bigg(1+\frac{\gamma_{W2}+\gamma_R+g}{\gamma_{W1}}\bigg)\\+\frac{\Gamma^4}{16\gamma^2_{W1}\gamma^2_{W2}r_m^2}\bigg].
\end{aligned}
\end{equation}

In the above, we assumed a lossless mirror, i.e. $t_m^2+r_m^2=1$. In the high gain limit, this expression reduces to $\bar{n}_{\ADD} = \frac{1}{2} + \frac{\gamma_{W2}}{\gamma_{W1}}=\frac{3}{2}|_{\gamma_{W1}=\gamma_{W2}}$, which is the same expression as the case with a perfect mirror.

\section*{Appendix C: Direct detection}
In the main text, we focused on noise arising when the homodyne detection scheme is used for the read-out of the optical signal. Here, we briefly discuss how these calculations will change if a direct detection scheme is instead used. In that case, the measured quantity is the light intensity rather than the quadratures of the electric field. As usual, the added noise is defined as the difference between the output noise and the amplifier input noise normalized by the gain value: $\text{Var}(\hat{S}^{\dagger}_{\OUT}\hat{S}_{\OUT})/|G|^2-\text{Var}(\hat{S}^{\dagger}_{\IN}\hat{S}_{\IN})$, where $\text{Var}(\hat{A})\equiv \braket{\hat{A}^2}-\braket{\hat{A}}^2$ for any operator $\hat{A}$. In terms of the noise operator $\hat{\mathcal{N}}$, direct detection with photon counting is given by the normal-ordered noise expression: $\bar{n}_{\ADD} \equiv \mathcal{S}[\omega_o]$, where 
$\mathcal{S} [\omega] \equiv \int d\omega'/2\pi \;  \braket{\hat{\mathcal{N}}^{\dagger}[\omega'] \hat{\mathcal{N}}[\omega]}$, which in turn reduces to:

\begin{equation} \label{Eq.Noise_Direct}
        \mathcal{S} [\omega] =\sum_{n}^{}|d_n [\omega]|^2.
\end{equation}

Note that, while the above expression depends directly on the coefficients of the noise terms associated with the gain channels, i.e. $d_n$, it has an indirect dependence on the loss channels in the high gain limit since these latter channels affect the lasing threshold. As an example, let us consider again the structure shown in Fig. \ref{Fig.Mirror_Amp}, which implements a chiral EP. From Eq. (\ref{Eq.SoutChiral}), we immediately see that under the approximation $\gamma_R \ll \gamma_{W1,2}$, the added noise in the high gain limit (obtained at the lasing threshold $g=\gamma_{W1}+\gamma_{W2}$) is given by $\bar{n}_{\text{add}}=g/\gamma_{W1}$, which becomes $\bar{n}_{\text{add}}=2$ for $\gamma_{W1}=\gamma_{W2}$. For comparison, we also note that this noise level is identical to that obtained for the DP amplifier of Fig. \ref{Fig.Linear_Amp}. For a PT amplifier with two waveguides (see Fig. \ref{Fig.PT_Amp}), we find that $\bar{n}_{\ADD}=g_1/\gamma_W$, which is large in the high gain limit. This behavior is similar to the noise associated with the homodyne detection scheme. From this discussion, it is evident that while changing the detection scheme affects the overall value of the noise, it does not have a significant impact on the ratio between noise levels in EP-based and DP-based optical amplifiers.

\begin{acknowledgments}
This project is supported by the Air Force Office of Scientific Research (AFOSR) Multidisciplinary University Research Initiative
(MURI) Award on Programmable systems with non-Hermitian quantum dynamics
(Award No. FA9550-21-1-0202). R.E. also acknowledges support from ARO (Grant No. W911NF-17-1-0481), NSF (Grant No. ECCS 1807552), and the Max Planck Institute for the Physics of Complex Systems. S.K.O. also acknowledges support from NSF (Grant No. ECCS 1807485). AM acknowledges funding by the Deutsche
Forschungsgemeinschaft through the Emmy Noether pro-
gram (Grant No. ME 4863/1-1) and the projects CRC
910 and CRC 183. AE acknowledges support from the DFG via a Heisenberg fellowship (Grant No EI 872/5-1).
\end{acknowledgments}

\bibliography{Reference}

\end{document}